\documentclass[%
preprint,
 amsmath,amssymb,
prl,
 longbibliography,
]{revtex4-1}
\usepackage{graphicx}
\usepackage{color}

\def\bd{\begin{displaymath}}
\def\be{\begin{equation}}
\def\ed{\end{displaymath}}
\def\ee{\end{equation}}

\begin{document}

\title{The Origin of the Spectral Intensities of Cosmic-Ray Positrons.}

\author{R. Cowsik, B. Burch and T. Madziwa-Nussinov}
\affiliation{\small { Physics Department and McDonnell Center for the Space Sciences\\
Washington University, St. Louis, MO 63130\\}}
\date{ \today \textup{ } } 
\begin{abstract}
\noindent We present a straightforward model of cosmic-ray propagation in the Galaxy that can account for the observed cosmic-ray positrons entirely as secondary products of cosmic-ray nucleons interacting with the interstellar medium. In addition to accounting for the observed energy dependence of the ratio of positrons to total electrons, this model can accommodate both the observed energy dependence of secondary to primary nuclei, like Boron/Carbon, and the observed bounds on the anisotropy of cosmic rays. This model also predicts the energy dependence of the positron fraction at energies higher than have been measured to date, with the ratio rising to $\sim$0.7 at very high energies. We briefly point out the differences between this model and the model currently in wide use that does not account for the observed positrons as secondaries and so prompts the interpretation of the observations as evidence for alternate origins of positrons.
\end{abstract}
\maketitle

\section{Introduction and Overview}

The recent measurement of the positron fraction $R_{e+}(E)=F_{e+}/(F_{e+}+F_{e-})$  at energies $E$  up to 300 GeV by the AMS collaboration \cite{Aguilar 2013} is an important contribution to cosmic-ray physics and poses a challenge to predict $F_{e+}(E)$ with similar precision. 
 
A striking feature of the AMS data confirming, with high statistics, earlier observations \cite{Adriani2009, Ackerman2012a} with unprecedented accuracy, is the monotonic \textit{increase} of $R_{e+}(E)$ from  $ \sim $0.052 at $ \sim $10 GeV to $ \sim $0.155 at $\sim$300 GeV. This contradicts an earlier calculation of $R_{e+}(E)$ expected assuming positrons are secondaries produced by primary cosmic-ray nuclei in the interstellar medium, by  Moskalenko and Strong (MS) \cite{Moskalenko1998} which predicted a monotonic \textit{decrease} (see the lower panel in  Fig. 1). Analysis with the MS model seems to require new physics to explain the discrepancy. It has been proposed that the positron excess could either originate from pulsars' magnetospheres \cite{ Yuksel2012} or from the annihilation or decay of dark matter \cite{ Bergstrom2009}, a more exciting explanation which is constrained by the absence of high energy gamma rays from the center of the Galaxy \cite{Ackerman2012b}. A kinematical cut-off in the positron spectrum at $E \sim(\frac{1}{2}-\frac{1}{4})$M(x), with M(x) being  the mass of the dark matter, is generally viewed as an indication for dark matter. 
 
We show here that the observed decrease of $R_{e+}(E)$ up to $\sim$6 GeV and its subsequent increase can be explained entirely as cosmic-ray secondaries, if positrons have a $1-2$ Myr residence time in the general interstellar medium, independent of their energies. Our calculations predict a dip in the positron fraction beyond 300 GeV followed by an increase beyond 1000 GeV to reach an asymptotic value of $\sim$0.7. This prediction is shown in the lower panel of Fig. 1. The difference of up to a factor $\sim$10 at 300 GeV between our calculated $F_{e+}(E)$ and that predicted by MS \cite{Moskalenko1998} indicates that present theoretical calculations are model-dependent and do not allow us to reach firm conclusions \cite{Cowsik2009, Mertsch2009, Shaviv2009, Katz2010, Cowsik2010, Gaggero2013}, including that decay or annihilation of massive dark matter in our Galaxy  is responsible for the positron spectrum.

The rest of this letter is devoted to a presentation of the analysis leading to our results and an explanation of the differences between the models for cosmic-ray propagation suggested by the positron observations and those currently in vogue for analyzing cosmic-ray observations. 
\begin{figure} [ht]
\includegraphics[width=11cm]{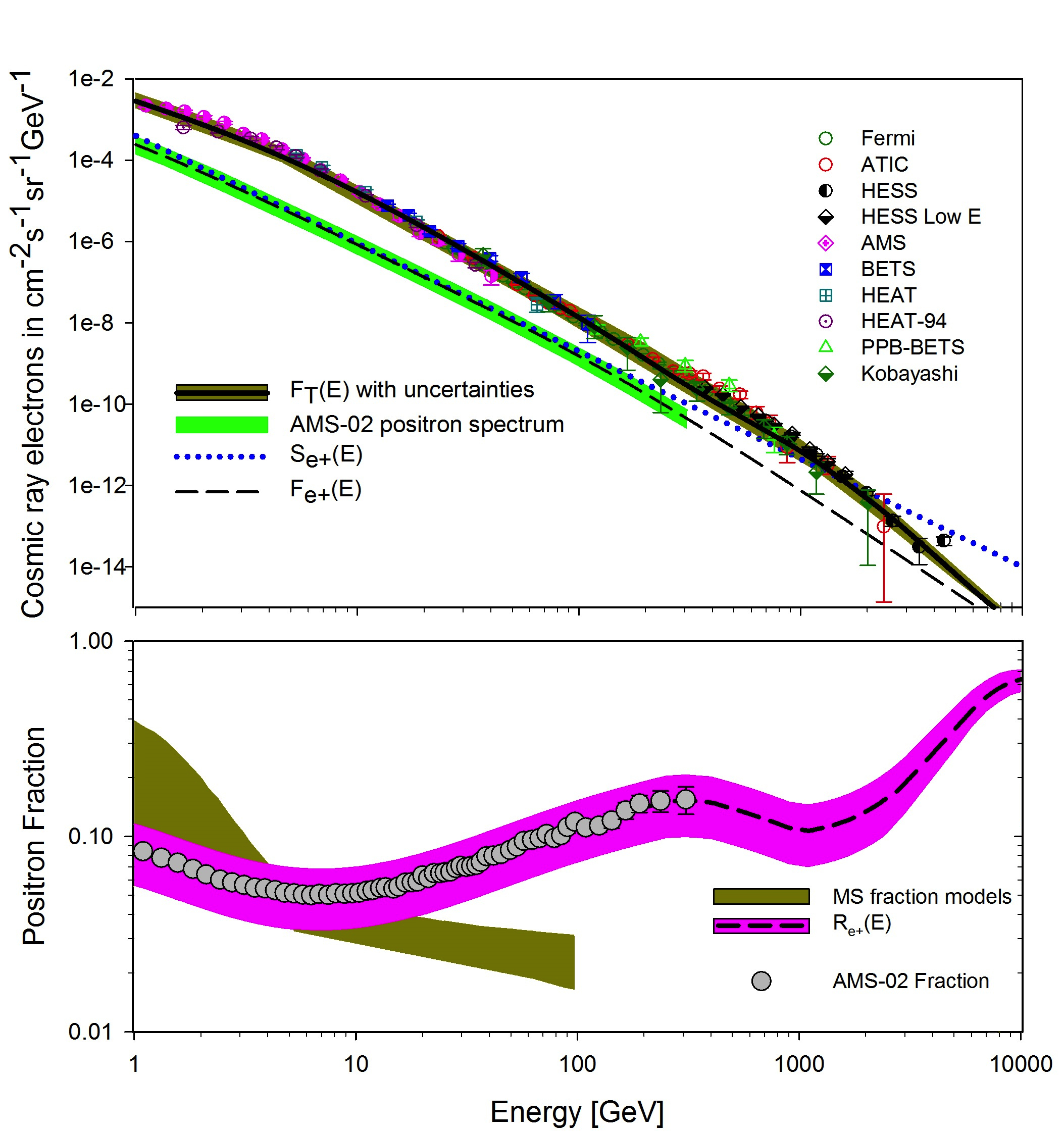} 
\caption{ Upper panel: The solid black line represents our fit, $F_{T}(E)$, to the spectrum of the total electronic component observed  in cosmic rays, the dotted line shows the source function $S_{e+}(E)$, with n$_{H} = 1$ cm$^{-3}$ and $\tau$ = 1 Myr, which fortuitously lies very close to the data points  representing the observed positron spectrum $F_{AMS}$ obtained by multiplying AMS-02 data on positron fraction by $F_{T}(E)$. The dashed line represents $F_{e+}(E)$ the theoretical spectrum with $S_{e+}(E)$ as the source function, including propagation effects during a residence time  $\tau$ = 2  Myr and  n$_{H} = 0.5$ cm$^{-3}$. Lower panel: Our predicted  positron fraction, $R_{e+}(E)=F_{e+}/ F_{T}$, with uncertainties is shown; the shaded steeply falling region is due to MS models \cite{Moskalenko1998}.}
\label{fig:Fig1}
\end{figure}

\section{Analysis of the Models}
The spectral intensities of cosmic-ray positrons provide a hitherto unavailable probe for the study of the origin and propagation of cosmic rays. First, they are not ubiquitous like  electrons, which may be accelerated to cosmic-ray energies in the sources; positrons have to be generated either as secondaries through the $\pi^{+} \rightarrow \mu^{+} \rightarrow e^{+}+\nu_{e} +\bar{\nu}_{\mu}$  and other decay chains of mesons produced in the cosmic-ray interactions in the interstellar medium or should be generated in exotic processes in pulsar magnetospheres or through the decay or annihilation of dark matter. Secondly, in high energy cosmic-ray proton interactions, the positrons carry away a small fraction $\sim$3-6$\%$ of the energy of the primary, unlike secondary nuclei, like B, which emerge with the same energy per nuclei as their parent nuclei, C, O, etc. Thus the observed spectrum of positrons will carry a signature of its origins. 

We have calculated the spectrum of positrons,  $F_{e+}(E)$,  by multiplying the positron fraction, $R_{e+}(E)$, measured by the AMS instrument by a fit to the spectrum of the total electronic component, $F_{T}(E)$. Both these spectra are displayed in Fig. 1. The PAMELA results \cite{Adriani2009} are consistent with $F_{e+}(E)$ displayed here. The spectrum of positrons $F_{e+}(E)$ has the form $A\cdot E^{-\beta _{+}}$ with $\beta_{+} \sim$ 2.65, almost identical with that of the total nuclear component of primary cosmic rays. In contrast, the total electronic component has a spectrum that has a spectral index $\beta_{T} \approx$ 2.2 below a few GeV, steepening to an index of $\beta_{T}\approx$ 3.1 until $\sim$1000 GeV, beyond which there is a rapid decrease of the intensities.

The rate of generation of positrons in the interstellar medium depends on the well-measured intensities of the nuclear component of primary cosmic rays and the cross sections for meson production, which exhibit a scaling behavior with adequate accuracy. A recent compilation of cosmic-ray proton and He spectra at high energies \cite{Bernard2012}, yields a spectrum for the nucleons with spectral index $\sim$2.65 in the relevant energy region. Based on this spectrum and the calculations of MS (see Fig. 4 in \cite{Moskalenko1998}), we estimate the source spectrum of the positrons to be
\begin{equation}
\label{eq:1}
S_{e+}(E) = S_{0}n_{H}E^{-\beta _{+}}
\end{equation}
 with S$_{0}\approx$ 4$\times$10$^{-4}$cm$^{-2}$s$^{-1}$sr$^{-1}$GeV$^{-1}$Myr$^{-1}$ and $\beta_{+} \approx$ 2.65, when n$_{H}$ is given in terms of the  number of hydrogen atoms cm$^{-3}$ and $E$ is in GeV. We display in Fig. 1, $S_{e+}(E)$ for n$_{H}$ = 1 cm$^{-3}$, which is a remarkably close fit to the observed spectrum in the energy band $3-100$ GeV, with a fortuitously close normalization. With n$_{H}$ = 1 cm$^{-3}$, 1 Myr corresponds to an effective grammage, $\Lambda_{e+}\approx$ 1.7 g$\cdot$cm$^{-2}$ for the positrons. In order to obtain the steady state spectrum of positrons, we should account for the energy loss suffered by positrons through synchrotron radiation and inverse Compton scattering against the 2.7 K microwave background in the Galaxy. This loss is assumed to be smooth and is parametrized as
\begin{equation}
\label{eq:2}
\frac{dE}{dt}=-bE^2
\end{equation}
where $b \approx$ 1.6$\times$10$^{-3}$GeV$^{-1}$Myr$^{-1}$ and $E$ is in GeV \cite{Cowsik2010}. 

The transport of cosmic-ray positrons and electrons may be described by the equation which includes the spatial diffusion, energy losses due to synchrotron emission and Compton scattering in the Galactic magnetic field and of the universal microwave background respectively, and the ultimate leakage from the Galaxy in an appropriate way: 
\begin{equation}
\frac{\partial n}{\partial t}-\nabla\cdot(\kappa\nabla n)-\frac{d}{dE}(bE^2n)-\frac{n}{\tau}=Q,
\end{equation}
where $Q$ represents the source term. It can be shown that the transport equation admits the Green's function
\begin{equation}
G(r,E,t)=(4\pi\kappa t)^{-3/2}\exp\bigg(-\frac{r^2}{4\kappa t}-\frac{t}{\tau}\bigg)(1-bEt)^2\delta\bigg(E_0-\frac{E}{1-bEt}\bigg).
\end{equation}
The represents the intensity of electrons or positrons seen at $r=0$, time $t$, and energy $E$ due to an impulse at $t=0$, position $r$, and energy $E_0$. This equation indicates that for a continuously emitting source at a distance $r$, admitting a power-law spectrum of electrons, there will be a sharp steepening of the spectrum beyond $E\gtrsim\frac{4\kappa}{br^2}$. This aspect may be applied to interpret the sharp steepening in the primary electron spectrum observed at $\sim1$ TeV.

For positrons which are produced as secondaries in the interstellar medium, the source function is continuous, uniform, and is a power law in energy. The observed spectrum is an integral over these distributions:
\begin{eqnarray}
\label{eq:3}
F_{e^+ISM}(E)&=&\int dt \int 4\pi r^2 dr\int(4\pi\kappa t)^{-3/2}e^{-\frac{r^2}{4\kappa t}-\frac{t}{\tau}}(1-bEt)^2\delta\bigg(E_0-\frac{E}{1-bEt}\bigg)\frac{S_0 n_H}{E_0^{\beta_+}}dE_0\nonumber\\
&=&\int_0^{1/bE}\frac{S_0}{E^{\beta_+}}(1-bEt)^{\beta_+ -2}e^{-t/\tau}dt\\
&\approx& S_0 n_{H}\tau E^{-2.65} \textup{ for}\textup{  } \textup{  }1 \textup{ GeV}< E \lesssim 300 \textup{ GeV} \nonumber \\
&\approx& \frac{S_0}{1.65b}n_{H}E^{-3.65} \textup{ for}\textup{  } E\gtrsim 300 \textup{ GeV.}\nonumber
\end{eqnarray}
similar to earlier results \cite{Cowsik1966}. As positrons enter the heliosphere and propagate to  near-Earth locations, they suffer solar modulation. 
Accounting approximately for these effects,  which affect only the very lowest end of their spectrum, we get
\begin{equation}
\label{eq:4}
F_{e+}(E) = F_{e+ISM}(E)e^{-E_{m}/E}
\end{equation}
 and display this spectrum in Fig. 1 for  n$_{H} $= 0.5 cm$^{-3}$, $\tau$ = 2 Myr and $E_m$ = 0.5 GeV, which provide a good fit to the data. The product   n$_{H} \tau \approx$ 1 corresponds to a grammage $\Lambda_{e+}\approx \textup{cm}_{H}\textup{n}_{H} \tau \approx 1.7 \textup{ g$\cdot$cm}^{-2}$. Note that the energy losses have steepened the high energy part of the positron spectrum to $E^{-3.65}$ and has made the spectral intensity independent of the leakage lifetime, a result that can be shown to be true even if $\tau$ were to be dependent on energy. Note that the calculated spectrum of secondary positrons fits the observations well. The free parameters in this fit are the leakage lifetime, $\tau \approx$ 2 Myr and the mean interstellar density  n$_{H}$ = 0.5 cm$^{-3}$. In the lower panel, the positron fraction,  $R_{e+}(E)$  is displayed and fits the observations, as expected, because the e$^+$ spectrum agrees well with the calculation.

What is the expected behavior of the positron fraction at higher energies? The answer to this question rests on the following considerations: In the energy region up to $\sim$2000 GeV, we have the observations of the total electronic component. Accordingly, the ratio $R_{e+}(E)= F_{e+}(E)/F_{T}(E)$  is easily calculated. The sharp steepening of $F_{T}(E)$ beyond 1000 GeV is attributed to the discrete nature of the cosmic-ray sources and the energy losses suffered by the electronic component in the finite amount of time needed for them to arrive at the Earth \cite{Cowsik1979, Cowsik2009, Cowsik2010, Shaviv2009, Nishimura1997}. Even though the primary electronic component cuts off, the secondary electrons and positrons that are produced in the interstellar medium that surrounds the Earth suffer only the aforementioned steepening and continue as $\sim$$E^{-3.65}$ at least up to 10$^4$ GeV. At these energies they dominate the flux and their ratio will be controlled by their production characteristics. Keeping in mind that there exists an excess of protons over neutrons (bound in nuclei) in the primary cosmic-ray beam and the fact that inelastic diffraction in the forward direction dominates the secondary cosmic-ray flux, we may note that e$^+$ is favored over e$^-$ in the production and the secondary e$^-$ is about $\sim$0.5 of the e$^+$ \cite{Moskalenko1998}. Thus we expect the  $R_{e+}(E)$ to reach $\sim$0.7 at E $\gg$ 10$^3$ GeV.

Alternatively, following Daniel and Stephens \cite{Daniel1970}, the muon charge ratio observed in the Earth's atmosphere yields $R_{e^+}$ of secondaries to be $\sim0.6$, not too different from the theoretical value.

\begin{figure} [ht]
\includegraphics[width=11cm]{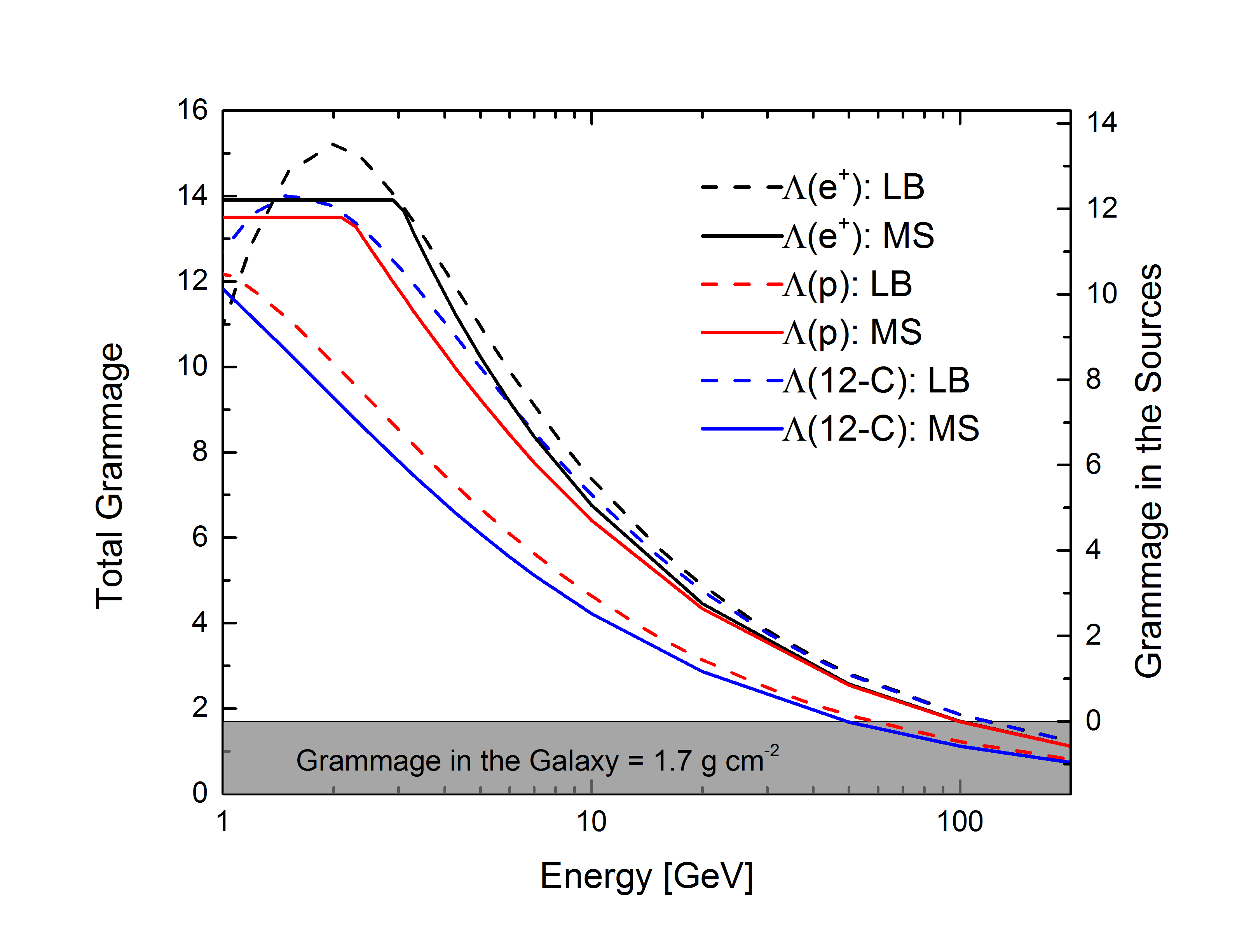} 
\caption{The grammage $\Lambda(E)$ for the various cosmic-ray particles as a function of the kinetic energy per nucleon or per positron is displayed for the leaky box model of Davis \textit{et al.} \cite{Davis2000}  as dashed lines and estimated from the diffusion model of MS \cite{Moskalenko1998}  as solid lines. In our model, much of the energy dependent part of the grammage is attributed to traversal in the sources and a constant value $\sim$ 1.7 g$\cdot$cm$^{-2}$, independent of energy, is traversed in the interstellar medium of the Galaxy. }
\label{fig:Fig2}
\end{figure}

\section{Discussion}

It would be appropriate now to address the question: what are the significant differences between the two models, one by MS  \cite{Moskalenko1998} and the other described here, both displayed in the lower panel of Fig. 1, that they make such diverse predictions for the positron fraction? The motivations for the cosmic-ray modeling has been provided by the observations of the ratio of secondary nuclei like B to that of their parent nuclei like C and O. Once the cross section per spallation is known then the grammage essentially controls the observed ratio. The effects of spallation and energy losses suffered by the secondary nuclei during their residence time in the Galaxy and traversing this grammage, \textit{on the average}, are to be taken into account \cite{Davis2000, Cowsik1967}.

The models currently in use are exemplified by those of Moskalenko and Strong  and of Davis  \textit{et al.} \cite{Moskalenko1998, Davis2000}. The  diffusion constant and other transport properties in the models are specified in terms of the rigidity (momentum per unit charge) and velocity. These are converted to grammage and are shown as a function of kinetic energy, $ E$, per nucleon or just kinetic energy for positrons in Fig. 2. These models predict essentially the same energy dependence at $ E$ $\gtrsim$ 1 GeV, but differ significantly at lower energies. However, at $\sim$3 GeV the grammage of positrons is $\sim$14 g$\cdot$cm$^{-2}$ and $\sim$7 g$\cdot$cm$^{-2}$ for Carbon because the Carbon nuclei have higher rigidity by a factor of A/Z and a lower velocity compared to the positrons. On the other hand, in the alternate model \cite{Cowsik2009, Cowsik2010} discussed here, all the particles are assumed to have the same grammage $\Lambda \sim$ 1.7  g$\cdot$cm$^{-2}$ \textit{in the Galaxy}. This is also shown in Fig.  2. The rest of the energy-dependent grammage left over at lower energies, needed to explain the B/C ratio, is attributed to traversal of material in the sources \cite{Cowsik2009, Cowsik2010}. 

The following points comparing and contrasting the two models are noteworthy: 

1. The energy dependence of $E^{-0.6}$  of current models \cite{Moskalenko1998} will steepen the production spectrum $S_{e+} \sim$ E$^{-2.65}$ to yield a steady state spectrum $E^{-3.65}$ and thence lead to a rapidly decreasing positron fraction R$_{e+}\sim$ $E^{-1}$, because the spectrum of the total electronic component is $\sim E^{-2.2}$ at $E < $ 6 GeV and $E^{-3.1}$ at higher energies.

2. If we normalize the current models with $\Lambda_{e+}(1\textup{ GeV} <E<3\textup{ GeV})\approx 14\textup{ g$\cdot$cm}^{-2}$ as expected by modeling the B/C ratio, then we \textit{over-produce} positrons by a factor of $\sim$7 at these energies.

3. This situation is to be contrasted with the model discussed here. The close similarity of the spectrum of positrons $E_{e+}(E)$ and those of the parent primary cosmic-ray nuclei allows a good fit to the observations of $R_{e+}(E)$ up to 300 GeV and predicts the expected behavior at high energies, reaching an asymptotic value of $\sim$0.7.

4. Over-production of positrons at low energies is avoided by the fact that the primary nuclei have $20-30$ times higher energy than the positrons they produce, i.e. they will be in the range $E > 20-30$ GeV where the residence time in the sources is so short that the parent nuclei leak out without significant positron production.

5. Finally, the $\sim$2 Myr residence time for all cosmic rays at least up to several hundred TeV allows one to predict anisotropies consistent with the observations (see \cite{Cowsik2009} for a compilation)  $\delta \approx $ 3$\kappa\triangledown$n/cn $ \approx$ 3r/4c$\tau \approx$ 5$\times$10$^{-4}$, for the length scale r $\approx$ 500 pc. As Strong, Moskalenko and Ptuskin have noted ( Fig. 12 in \cite{Strong2007}), the increase of the diffusion constant $\kappa $ as $E^{0.6}$, which yields  $\Lambda \sim$ $E^{-0.6}$, concomitantly generates anisotropies that increase with increasing energy. This has led to considerable tension between upper bounds on anisotropy and that predicted by current models of cosmic ray propagation.

\section{Summary}

In summary, we may state that, the production of positrons by nuclear primary cosmic rays interacting with the  interstellar medium provides a good explanation of the observed spectrum and ratio with the total electronic component, provided these particles have a residence time of $\sim$2 Myr in the interstellar medium, independent of their energy. 
The prediction of the positron fraction 
rests on a calculation of energy loss suffered by the positrons during their residence  for $\sim$2 Myr in the interstellar medium, which leads to $\sim E^{-3.65}$ for $E>$ 300 GeV. The total electronic component has a spectrum $\sim E^{-3.1}$ up to $\sim$ 1000 GeV and rapidly decreases in intensity at higher energies, so that the positron fraction tends to reach an asymptotic value dictated by their production characteristics as secondaries, $\sim$0.7 for $E \gg$ 1000 GeV. Because the sources of primary cosmic-ray electrons are discrete, their spectrum shows a cut-off at an energy $\sim$1000 GeV, dictated by the distance to the nearest source. The spectrum of primary electrons may be understood as the sum of the contributions from various discrete sources, with the nearest source dominating at the highest energies \cite{Cowsik2009, Cowsik2010, Cowsik1979, Nishimura1997}.

\section{Acknowledgement.}
We would like to thank professors M. H. Israel and S. Nussinov who played a key role in shaping these comments through discussions of all the key points made in the paper, and honing each scientific point to achieve clarity in the presentation.

\end{document}